\documentclass[notitlepage,prd,twocolumn,showpacs,preprintnumbers,nofootinbib,superscriptaddress,floatfix,tightenlines]{revtex4-2}

\usepackage{mathrsfs}
\usepackage{amssymb}
\usepackage{amsmath}
\usepackage{graphicx}
\usepackage{amsfonts,amsbsy}
\usepackage{bm}
\usepackage{nicefrac}
\usepackage{xcolor}
\usepackage[breaklinks,colorlinks,urlcolor=blue,citecolor=citcolor,linkcolor=lcolor]{hyperref}
\usepackage{cleveref}
\usepackage{xcolor}

\definecolor{lcolor}{rgb}{0.5,0,0}
\definecolor{citcolor}{rgb}{0,0.3,0.0}
\definecolor{ao(english)}{rgb}{0.0, 0.5, 0.0}
\usepackage{comment}

\newcommand{\pC}{p_{\rm L}}
\newcommand{\pQ}{p_{\rm H}}
\newcommand{\nC}{n_{\rm L}}
\newcommand{\nQ}{n_{\rm H}}
\newcommand{\muC}{\mu_{\rm L}}
\newcommand{\muQ}{\mu_{\rm H}}
\newcommand{\muB}{{\mu}}
\newcommand{\muc}{{\mu_{\rm c}}}
\newcommand{\cslim}{{c_{s,{\rm lim}}^2}}

\newcommand{\csminlim}{(c^2_s)^{\rm min}_{\rm lim}}

\begin{document}

\title{How perturbative QCD constrains the Equation of State at Neutron-Star densities}

\author{Oleg Komoltsev}
\affiliation{Faculty of Science and Technology, University of Stavanger, 4036 Stavanger, Norway}
\author{Aleksi Kurkela}
\affiliation{Faculty of Science and Technology, University of Stavanger, 4036 Stavanger, Norway}
\begin{abstract}
 We demonstrate in a general and analytic way how high-density information about the equation of state (EoS) of strongly interacting matter obtained using perturbative Quantum Chromodynamics (pQCD) constrains the same EoS at densities reachable in physical neutron stars. Our approach is based on utilizing the full information of the thermodynamic potentials at the high-density limit together with thermodynamic stability and causality.
This requires considering the pressure as a function of chemical potential $p(\mu)$ instead of the commonly used pressure as a function of energy density $p(\epsilon)$. The results can be used to propagate the pQCD calculations reliable around 40$n_s$ to lower
densities in the most conservative way possible. We constrain the EoS starting from only few times the nuclear saturation density $n \gtrsim 2.2 n_s$ and at $n = 5 n_s$ we exclude at least 65\% of otherwise allowed area in the $\epsilon - p$ -plane. This provides information complementary to astrophysical observations that should be taken into account in any complete statistical inference study of the EoS. These purely theoretical results are independent of astrophysical neutron-star input, and hence, they can also be used to test theories of modified gravity and BSM physics in neutron stars. 
\end{abstract}

\maketitle
\section{Introduction}
\label{sec0}
The rapid evolution of neutron-star (NS) astronomy --- in particular, the recent NS radius measurements \cite{Miller:2021qha,Riley:2021pdl}, the discovery of massive NSs \cite{Demorest:2010bx, Antoniadis:2013pzd,Fonseca:2021wxt}, and the advent of gravitational-wave and multi-messenger astronomy \cite{TheLIGOScientific:2017qsa,GBM:2017lvd} --- is for the first time giving us empirical access to the physics of the cores of NSs.  Within the Standard Model and assuming general relativity, the internal structure of NSs is determined by the equation of state (EoS) of strongly interacting matter  \cite{1939PhRv...55..364T, Oppenheimer:1939ne}.
With these assumptions, NS observations can be used to empirically determine the EoS \cite{Annala:2017llu,Margalit:2017dij,Rezzolla:2017aly,Ruiz:2017due,Bauswein:2017vtn,Radice:2017lry,Most:2018hfd,Dietrich:2020efo,Capano:2019eae,Landry:2018prl,Raithel:2018ncd,Raithel:2019ejc,Raaijmakers:2019dks,Essick:2019ldf,Al-Mamun:2020vzu,Essick:2021kjb,Paschalidis:2017qmb,Annala:2019puf,Ferreira:2020kvu,Minamikawa:2020jfj,Blacker:2020nlq} (for reviews, see~\cite{Baym:2017whm,Gandolfi:2019zpj,Raithel:2019uzi,Horowitz:2019piw,Baiotti:2019sew,Chatziioannou:2020pqz,Radice:2020ddv}). And if the EoS can be determined theoretically to a sufficient accuracy, comparison with NS observations allows to use these extreme objects as laboratory for physics beyond the standard model (e.g.,~\cite{Goldman:1989nd, Giudice:2016zpa,Ciarcelluti:2010ji, Li:2012ii, Xiang:2013xwa, Tolos:2015qra, Ellis:2018bkr, DelPopolo:2020hel, Jimenez:2021nmr}) and/or general relativity (e.g.,~\cite{Damour:1993hw,He:2014yqa,AparicioResco:2016xcm, Doneva:2018ouu,LopeOter:2019pcq}). 

For both of these goals, it is crucial to make use of all possible controlled theory calculations that inform us about the EoS at densities reached in NSs. 
While in principle the EoS is determined by the underlying theory of strong interactions, Quantum Chromodynamics (QCD), in practice we have access to the EoS only in limiting cases. In the context of low temperatures relevant for neutron stars, the EoS of QCD can be systematically approximated at low- and at high-density limits. At low densities, the current state-of-the-art low-energy effective theory calculations allow to describe matter to densities around and slightly above nuclear saturation density $n \approx n_s = 0.16/\textrm{fm}^3$ \cite{Tews:2012fj, Drischler:2017wtt}, but become unreliable at higher densities $n\sim 5-10 n_s$ reached in the cores of massive neutron stars. A complementary description of NS matter comes from perturbative QCD (pQCD) calculations which become reliable at sufficiently high densities $\sim 40n_s$, far exceeding those realised in NSs \cite{Gorda:2018gpy,Gorda:2021znl}.  

\begin{figure}[t]
    \centering
\includegraphics[width=0.5\textwidth]{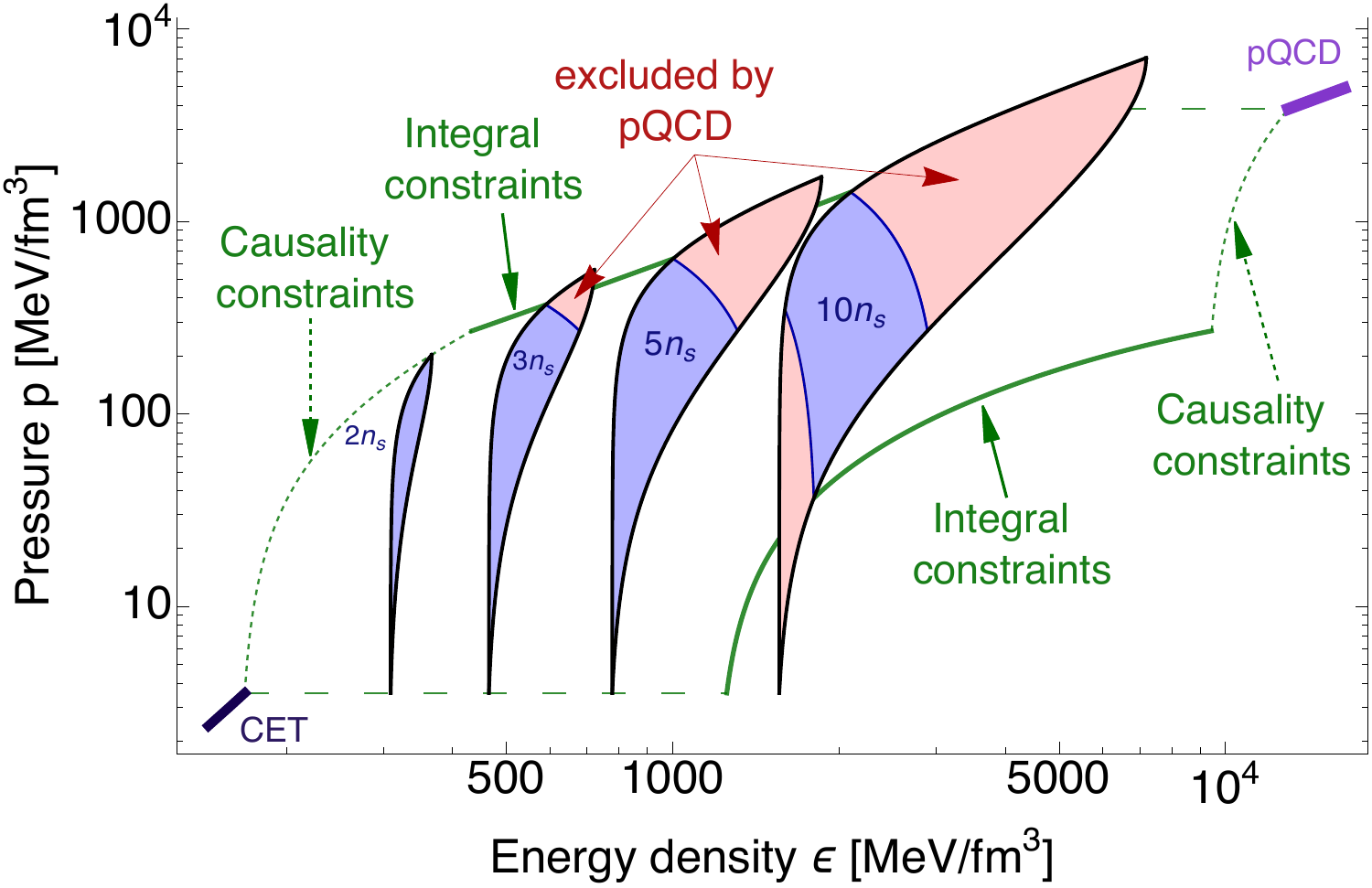}
    \caption{
    Theoretical constraints to $\epsilon$ and $p$ arising from low- (CET) and high-density input (pQCD). The green lines show the envelope of allowed values of $p(\epsilon)$; dashed green lines arise from trivially imposing causality of the EoS while the thick solid lines correspond to newly identified integral constraints arising from imposing simultaneously the high-density limit for $p$, $n$, and $\muB$. 
    The blue regions correspond to allowed values $\epsilon$ and $p$ at fixed density $n$; the red regions would be otherwise allowed but are excluded by the high-density input. At $n=5 n_s$, $75\%$ of otherwise allowed values of $\epsilon$ and $p$ are excluded by the high-density input.
    }
    \label{fig1}
\end{figure}

Several works have used large ensembles of parameterized EoSs to study the possible behavior of the EoS in intermediate densities between the theoretically known limits. Many of these works have anchored their EoS to the low-density limit only, while others have interpolated between the two orders of magnitude in density separating the low- and high-density limits  \cite{Kurkela:2014vha,Annala:2017llu,Most:2018hfd,Annala:2019puf,Annala:2021gom}. While the input from pQCD clearly constrains the EoS at very high densities, how this information affects the EoS around neutron-star densities has so far been convoluted by the specific choices of interpolation functions.

The aim of the present work is to make the influence of the high-density calculations to the EoS at intermediate densities explicit, and in particular, to derive constraints to the EoS
that are completely independent of any specific interpolation function.
By using the full information available in the thermodynamic grand canonical potential as a function of baryon chemical potential, $\Omega(\muB) = -p(\muB)$, we find stricter bounds than works using only the reduced form of the EoS, i.e.,~the pressure as a function of the energy density $p(\epsilon)$ appearing in the hydrodynamic description of neutron-star matter \cite{Rhoades:1974fn,Koranda:1996jm,LopeOter:2019pcq,Tews:2019cap, Lope-Oter:2021mjp}. The effect is demonstrated in \cref{fig1}, which shows the region in $\epsilon - p $ -plane that can be reached with a causal and thermodynamically stable EoS with information of the high-density limit, and in particular compares the allowed values at different fixed densities with and without the pQCD input.

\section{setup}
\label{sec1}

In the following we consider all possible interpolations of the full thermodynamic potential at zero temperature and in $\beta$-equilibrium,  $\Omega(\muB)=-p(\muB)$, between the low-density limit $\muB=\muC$ and the high-density limit $\muB=\muQ$. We assume that at both of these limits the pressure $p(\muB)$ and its first derivative, baryon number density $n(\muB) = \partial_{\muB}p(\muB)$ are now
\begin{align}
p(\muC) &= \pC,  \quad  p(\muQ) = \pQ, \\
n(\muC) &= \nC, \quad n(\muQ) = \nQ.
\end{align}
This information is readily available from the microscopic calculations which are assumed to be reliable at these limits; representative values from state-of-the-art chiral effective theory (CET) \cite{Hebeler:2013nza} and pQCD calculations \cite{Gorda:2021znl} from the literature are reproduced in \cref{tab1} \footnote{In addition to the perturbative contribution, the pressure at high densities may also receive a non-pertrubative, density independent contribution often referred as the \emph{bag constant.} We have checked that inclusion of a bag constant of $200$MeV/fm$^3$ leads to an effect that is qualitatively similar but negligible in size compared to scale variation error.}.

Thermodynamic consistency requires that pressure is a continuous function of $\muB$. Similarly, thermodynamic stability requires concavity of thermodynamic potentials $\partial_\muB^2 \Omega \leq 0$ so that $n(\muB)$ is a monotonically increasing function $\partial_\muB n(\muB) \geq 0$. Density $n(\muB)$ does not need to be continuous and it can have discontinuities (increasing the density at fixed $\muB$) in the case of first-order phase transitions; we place no additional assumptions on the number or strength of possible transitions. 
Causality requires that the speed of sound is less than the speed of light,  $c_s^2 \leq 1$, and it imposes a condition on the first derivative of $n$.
\begin{align} \label{eq3}
c_s^{-2} = \frac{\mu}{n}\frac{\partial n}{\partial \muB} \geq 1.
\end{align}

\begin{table}[h!]
    \centering
    \begin{tabular}{l|cc|ccc|}
        &\multicolumn{2}{c}{CET}&\multicolumn{3}{|c|}{pQCD}\\
        &soft&stiff&$X=1$&$X=2$&$X=4$\\
        \hline
        $\mu$ [GeV] & 0.966 & 0.978 & \multicolumn{3}{|c|}{2.6} \\
        $n$ [1/fm$^{3}$]& \multicolumn{2}{|c|}{0.176} & 6.14& 6.47&6.87\\
        $p$ [MeV/fm$^3$]& 2.163&3.542&2334.&3823.&4284.\\
        
    \end{tabular}
    \caption{Collection of predictions for the thermodynamic quantities at the low- (CET) and the high- (pQCD) density limits. The CET limit corresponds to the "stiff" and "soft" EoS of \cite{Hebeler:2013nza} while the pQCD values are from a partial N3LO calculation \cite{Gorda:2021znl} at three different values of the renormalization scale parameter $X$. The uncertainty of the pQCD limit can be assessed by varying $X$ in the range $X=[1,4]$%
    . All figures correspond to the "stiff" CET and $X=2$ pQCD. \Cref{figA1} in the Supplemental Material demonstrates the fairly mild $X$ dependence of the results. The value $\mu = 2.6$ GeV for pQCD is chosen so that the uncertainty in pQCD is roughly the same size as the uncertainty of CET at $n = 0.176$/fm$^3$. \cite{Kurkela:2014vha}}
    \label{tab1}
\end{table}

\begin{figure}[t!]
    \centering
\includegraphics[width=0.4\textwidth]{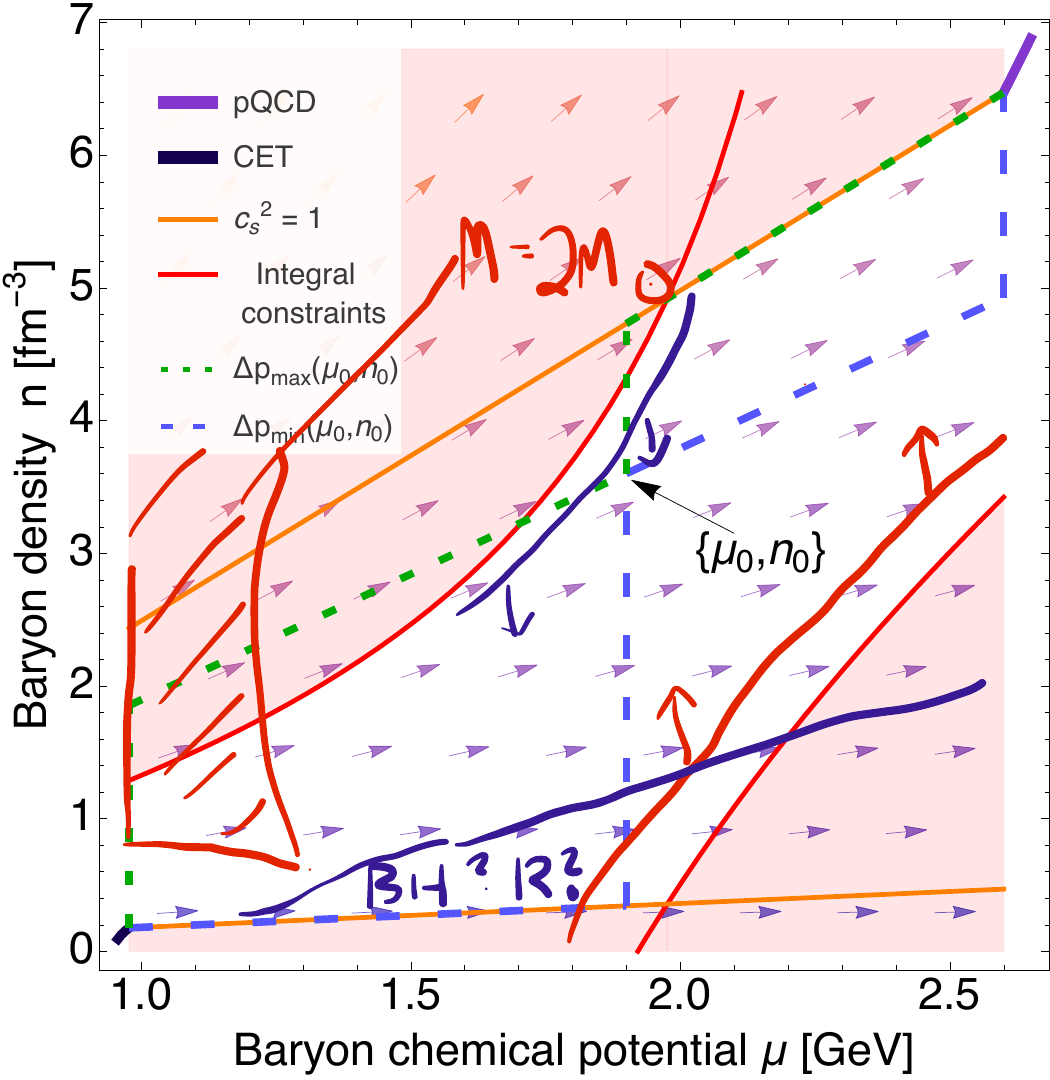}
    \caption{Density as a function of the chemical potential determines the EoS. At any given point $\{ \muB , n \}$, causality imposes a minimal slope to any EoS passing that point $\partial_\mu n(\mu) \geq \frac{n}{\mu}$, denoted by the arrows. Thermodynamic consistency demands that $n(\mu)$ is a monotonic function, and that the area under the curve in the plot is given by $\Delta p$. These conditions cannot be simultaneously fulfilled by any EoS that passes to the red area at any density and are therefore excluded. The lines $\Delta p_{\rm min}$ and $\Delta p_{\rm max}$ display the constructions defined in \cref{eq:n1} and \cref{eq:n2}.}
    \label{fig2}
\end{figure}

We construct the allowed EoSs by considering all possible functions $n(\muB)$ allowed by the above assumptions connecting the low- and high-density limits. 
Causality imposes a minimal slope, $\partial_\muB n(\muB) \geq n/\muB$, that any causal EoS passing though a given point in the $\muB - n$ plane can have. This is visualized as a vector field, where the arrows at each point correspond to tangent lines with constant $c^2_s=1$. This requirement imposes two fundamental constraints. 
Starting from the point $\{\muC , \nC\}$ we can follow the arrows until $\muQ$ by solving \cref{eq3} with $c^2_s = 1$, leading to $n(\muB) =\nC \muB/\muC $. This produces a maximally stiff causal EoS and the area under this line cannot be reached from the low-density limit with a causal EoS.
Correspondingly, the upper limit for the $n(\muB)$ can be obtained starting from $\{\muQ , \nQ\}$ and following the arrows backward to $\muC$; the high-density limit $\nQ$ cannot be reached from any point above this line by a causal EoS. These previously known bounds (e.g. \cite{Rhoades:1974fn,LopeOter:2019pcq}) are represented as orange lines in \cref{fig2}.

The simultaneous requirement of reaching both $\pQ$ and $\nQ$ imposes further constraints and fixes the area under the curve $n(\muB)$
\begin{align}
\int_{\muC}^{\muQ} n(\mu) d\mu= \pQ-\pC = \Delta p.
\end{align}
In the example shown in \cref{fig2}, this requirement imposes that the area under any allowed EoS is approximately one third the area of the figure. 

At each point, we can evaluate the absolute minimum and maximum area under any EoS ($\Delta p_{\rm min/max}$) that can be reached at $\muQ$ if the EoS goes through that particular point. If $\Delta p_{\rm min}$ ($\Delta p_{\rm max}$) is bigger (smaller) than $ \Delta p$, then such a point would be ruled out. 

To obtain the minimum area at $\muQ$ for any EoS going through a specific point $\{\mu_0, n_0\}$ consider the following construction shown in \cref{fig2} as a dashed blue line:
\begin{align}
\label{eq:n1}
    n(\muB) = 
    \left\{
    \begin{array}{rl}
    \nC \muB/\muC, & \muC < \mu < \mu_0 \\
    n_0 \muB/\mu_0, & \mu_0 < \mu < \muQ.
    \end{array}
    \right. 
\end{align}
For $\mu < \mu_0$, the smallest possible area is determined by the maximally stiff causal line (i.e.~$c_s^2 =1$) starting from $\{\muC ,\nC\}$ which we follow up to $\mu_0$. At $\mu_0$, we have a phase transition where the density jumps to $n_0$. After that, the EoS follows the maximally stiff causal line starting at $\{n_0,\mu_0\}$ until $\muQ$, where the EoS has another phase transition to reach $n(\muQ)=\nQ$. The solution to the equation $\Delta p_{\rm min}(\mu_0,n_0) = \int_{\muC}^{\muQ} n(\muB) = \Delta p $ is shown in \cref{fig2} as the top red line denoted integral constraints; any EoS crossing this line is inconsistent with simultaneous constraint on $\pQ$ and $\nQ$. This yields a maximum density for given $\muB$,
\begin{align}
\label{eq:nmax}
    n_{\rm max}(\muB) = 
    \left\{
    \begin{array}{ll}
     \frac{\muB^3 \nC - 
 \muB \muC (\muC \nC + 2 \Delta p)}{(\muB^2 - \muQ^2) \muC },
    &\muC \leq \muB < \muc \\
   \nQ \muB/\muQ, & \muc \leq \muB \leq \muQ,
    \end{array}
    \right. 
\end{align}
where $\muc$ is given by the intercept of the causal line and the integral constraint, i.e., the two cases in \cref{eq:nmax},
\begin{equation}
\label{eq:muc}
\muc = \sqrt{\frac{\muC \muQ ( \muQ \nQ-  \muC\nC-2 \Delta p )}{\muC \nQ - \muQ \nC} }.
\end{equation}

Similarly, the procedure to maximize area under any EoS going through the point $\{n_0,\mu_0\}$ is shown as a dashed green line in \cref{fig2},
\begin{align}
\label{eq:n2}
    n(\muB) = 
    \left\{
    \begin{array}{rl}
    n_0 \muB/\mu_0, & \muC < \muB < \mu_0 \\
    \nQ \muB/\muQ, & \mu_0 < \muB < \muQ.
    \end{array}
    \right.
\end{align}
Correspondingly, solving for $\Delta p_{\rm max} = \Delta p$ gives a constraint to the minimal $n$ that can be obtained for a given $\muB$, depicted as the bottom red line in \cref{fig2}. 
Then, for a given chemical potential the minimal allowed density is
\begin{align}
\label{eq:nmin}
    n_{\rm min}(\muB) = 
    \left\{
    \begin{array}{ll}
    \nC \muB/\muC, & \muC \leq \muB \leq \muc \\
   \frac{\muB^3 \nQ - 
 \muB \muQ (\muQ \nQ - 2 \Delta p)}{(\muB^2 - \muC^2) \muQ }, & \muc < \muB \leq \muQ.
    \end{array}
    \right. 
\end{align}

Note that the point of interception $\muc$ is the same for the upper and lower limits. This happens because the EoS following $n_{\rm min}(\muB)$ up to $\muc$ obtains the correct area $\Delta p$ only if the EoS jumps at $\muc$ from $n_{\rm min}(\muc)$ to $n_{\rm max}(\muc)$. This is also the EoS exhibiting a phase transition with the largest possible latent heat 
\begin{align}
    Q = \mu_c ( n_{\rm max}(\muc) - n_{\rm min}(\muc)) = \notag \\
    -\muC \nC + \muQ \nQ + 2 \pC - 2 \pQ.
\end{align} 

Further, we denote the special EoS with $c_s^2 = 1$ throughout the whole region as
\begin{equation}
\label{eq11}
n_c(\muB) = n_{\rm max}(\muC) \muB/\muC  = n_{\rm min}(\muQ) \muB/\muQ.
\end{equation}
Any point along this line maximises the area to the left of the point ($\muB < \mu_0$) and minimizes to the right of the point  ($\muB > \mu_0$), so that this line corresponds to the maximal pressure at fixed $\muB$. 

\section{Mapping to the $\epsilon - p$ plane}
\label{sec2}

\begin{figure}
    \centering
\includegraphics[width=0.5\textwidth]{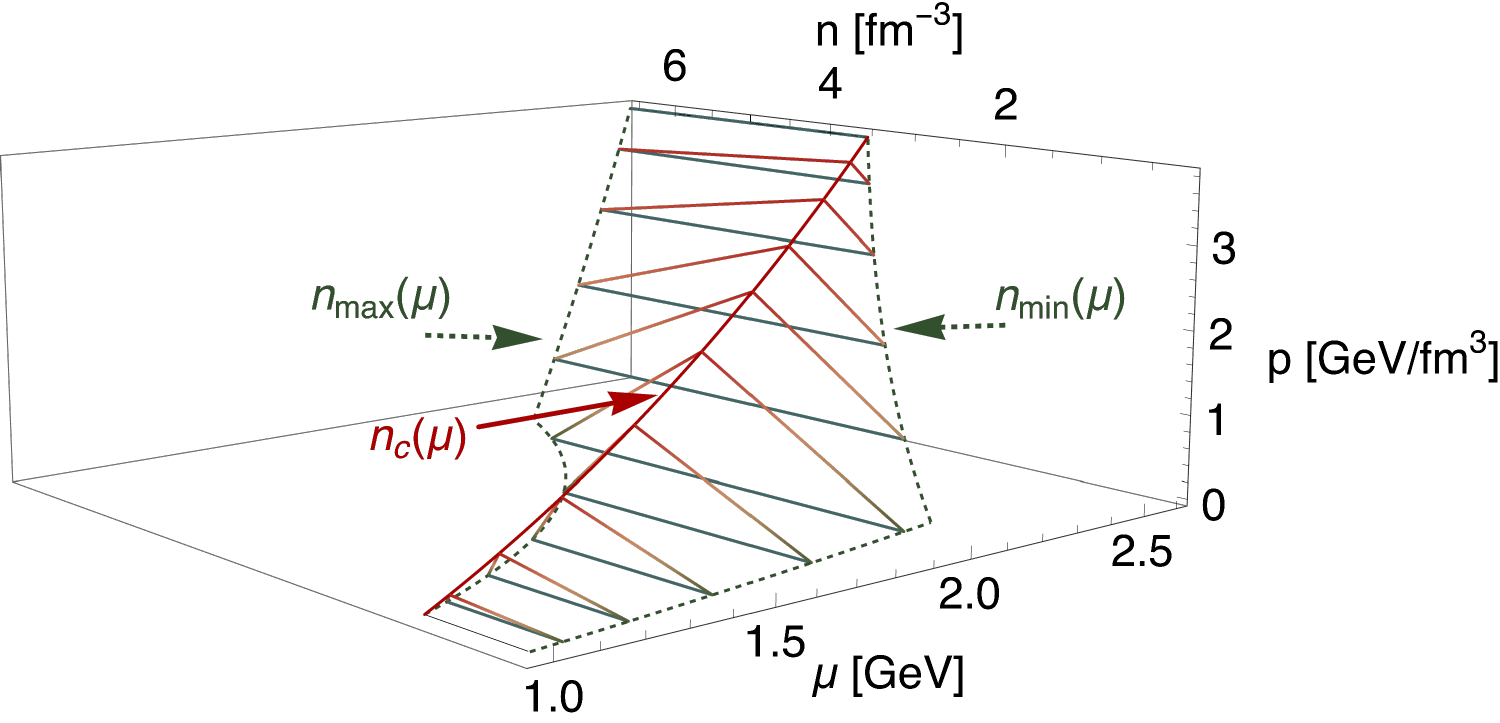}
    \caption{3D rendering of the constraints in the $\muB - n - p$ phase space. Each triangle is a slice of the $p-n$ constraints for fixed $\muB$. Note that $n_{max}, n_{min}$ and $n_c$, which are defined in \cref{eq:nmax}, \cref{eq:nmin}, and \cref{eq11}, respectively, are projection on the $\mu - n$ -plane of the lines denoted with the same names in the figure above.}
    \label{fig3}
\end{figure}

For every allowed point on the $\mu - n$ -plane, we can find minimal and maximal pressures, $p_{\rm min/max}(\mu_0,n_0)$, that can be obtained at that point $\{\mu_0 , n_0\}$. Note that this is different from the minimal and maximal pressures ($\Delta p_{\rm min/max}$) obtainable at $\muQ$ by an EoS passing though $\{\mu_0 , n_0\}$, as discussed in the previous section. 

The minimal pressure is given by the EoS that follows the maximally stiff causal line terminating at the point $\{\mu_0 , n_{\rm min}(\mu_0)\}$, i.e., $n(\muB) = \frac{\muB}{\mu_0} n_{\rm min}(\mu_0)$, and has a phase transition from $n_{\min}$ to $n_0$ at $\mu_0$.
This construction leads to a lower bound of the pressure as function of $\muB$ and $n$
\begin{align}
    p_{\rm min}(\mu_0,n_0) &
     = \pC + \frac{\mu_0^2-\muC^2}{2 \mu_0}n_{\rm min}(\mu_0) 
    \label{eq:pmin}
\end{align}

In order to find the maximal pressure for a given point, the $\mu - n$ plane needs to be divided in two different regions. For $n < n_c(\muB)$ (see \cref{eq11}), the maximal pressure is obtained by following the maximally stiff causal line terminating at the point itself $n(\muB) = n_0 \muB/\mu_0 $ 
\begin{align}
    p_{\rm max}(\mu_0,n_0) 
    &= \pC + \frac{\mu_0^2-\muC^2}{2 \mu_0}n_0, \quad n < n_c(\muB).
    \label{eq:pmax1}
\end{align}

For $n > n_c(\mu)$, the above construction would lead to a phase transition at CET point that is inconsistent with the upper integral constraints. Instead, a bound can be obtained by noting that the EoS that maximizes the pressure at $\{\mu_0,n_0\}$ is an EoS that minimizes the pressure difference between $\mu_0$ and $\muQ$. Thus the maximum pressure at that point is given by the difference between $\Delta p$ and the pressure for the maximally stiff EoS following the causal line starting at $\{\mu_0, n_0\}$
\begin{align}
    p_{\rm max}(\mu_0,n_0)
    &= \pQ - \frac{\muQ^2-\mu_0^2}{2 \mu_0}n_0,\quad n > n_c(\muB).
    \label{eq:pmax2}
\end{align}

The simultaneous bounds for $\muB$, $n$ and $p$ are visualized in \cref{fig3}. These constraints can be easily translated into bounds on the $\epsilon - p$ -plane using the Euler equation $\epsilon = -p + \muB n$ for a fixed density $n$. An analytic solution for the envelope of the allowed values of $p(\epsilon)$ irrespective of $n$ (green lines in \cref{fig1} and \cref{fig4}) is given in supplemental material.

\Cref{fig1} shows the allowed range of $\epsilon$ and $p$ values with and without imposing the high-density constraint for fixed densities of $n=2, 3,5,$ and $10 n_s$ using the standard central renormalization scale commonly used in the literature \cite{Schneider:2003uz,Ipp:2003jy,Fraga:2004gz}, $X=2$, for pQCD and the "stiff" values for CET. At the density of $n=2 n_s$, the high-density input does not offer additional constraint. However, strikingly, already at $n = 3 n_s$ the largest pressures are cut off by the integral constraint and only 68$\%$ of the $\epsilon -p $ values remain (given by the ratio of the blue and combined blue and red areas in \cref{fig1} when plotted in linear scale). At $n = 5 n_s$ only $25\%$ of the allowed region remains whereas at $n=10 n_s$,
the allowed range of values is significantly reduced, now also featuring a cut of the lowest pressures leaving around 6.5$\%$ of the total area.

We have checked the stability of these results 
against the variation of different pQCD and CET limits in \cref{tab1}. While varying the CET parameters has a very small effect on the excluded areas (of the order of line width in \cref{fig1}), varying the pQCD renormalization parameter increases the area in $\epsilon- p$ -plane. The effect of varying $X=[1,4]$ is shown in \cref{figA1} of the Supplemental Material. The union of allowed areas in range $X=[1,4]$ excludes $[13\%,64\%,92.8\%]$ of the otherwise allowed area (without pQCD) for  $n = [3,5,10]\ n_s$. The lowest density at which $p$ and $\epsilon$ are limited by the high-density input from above is $n_{\rm min}(\muC) = [2.2,2.5]\, n_s $ and from below is $n_{\rm max}(\muC) = [4.8,9.0] \, n_s $ for $X=[1,4]$.

\section{Speed-of-sound constraints}
The EoSs that render the boundaries of the allowed regions in \cref{fig1} and \cref{fig2} are composed of the most extreme ones. They contain maximal allowed phase transitions and extended density ranges where $c^2_s = 1$. While it is clear that these most extreme EoSs are unlikely to be the physical one, they cannot be excluded with the same level of robustness as those which break the above criteria. 

A possible way to quantify how extreme the EoSs are is by imposing a maximal speed of sound $c_s^2 < \cslim < 1$ that is reached at any density within the interpolation region \cite{Annala:2019puf}. In supplemental material we give the generalizations 
for $n_{\rm min/max}$, $p_{\rm min/max}$, and $\muc$ for arbitrary limiting $\cslim$.

\begin{figure}
    \centering
    \includegraphics[width=0.45\textwidth]{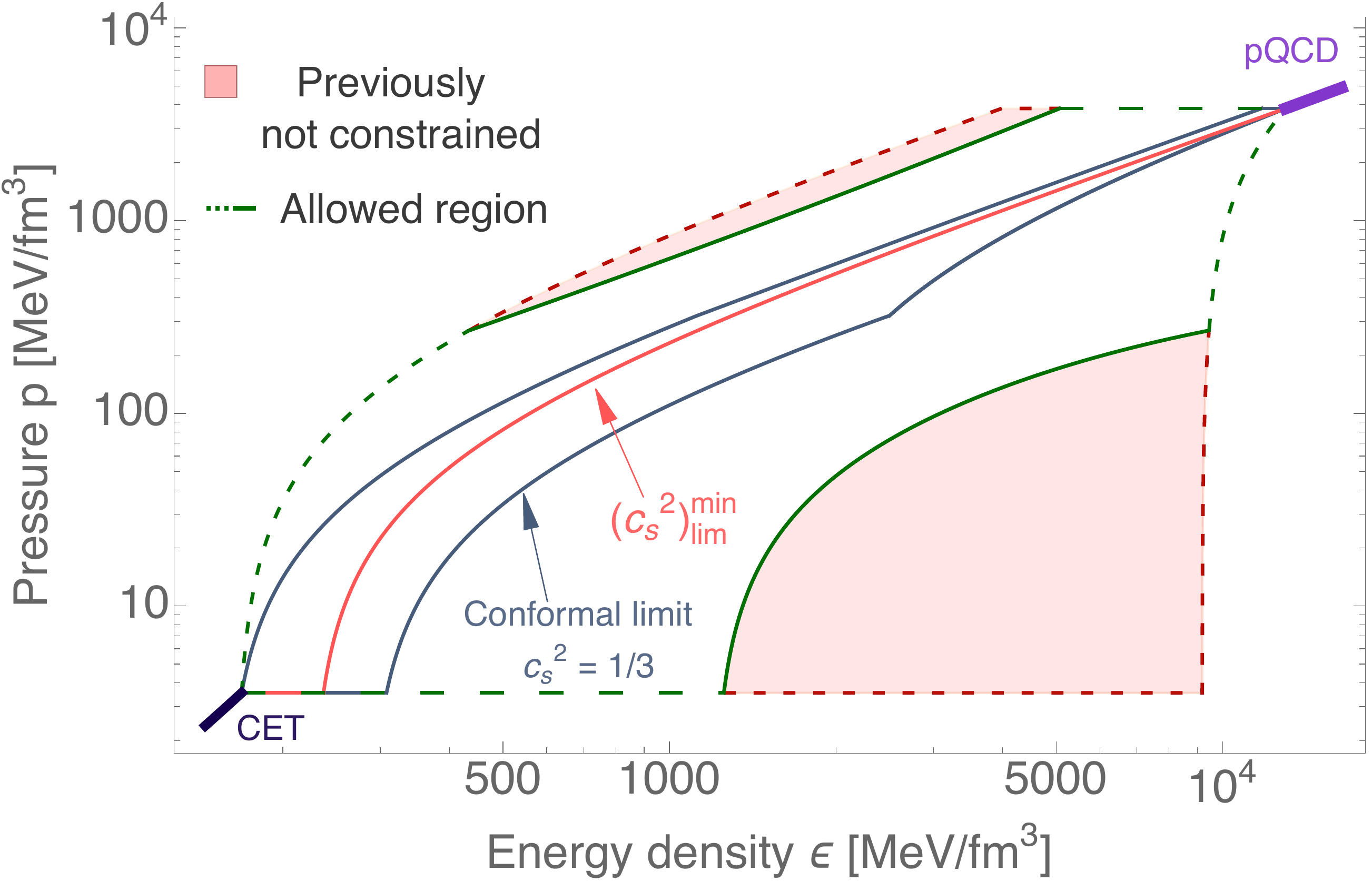}
    \caption{Constraints to $p(\epsilon)$ for different limiting values of the sound speed. The green envelope corresponds to $c_s^2 = 1$ while the blue envelope corresponds to the conformal limit $c_s^2 = 1/3$. At the specific value of the sound speed $\csminlim$ the allowed region degenerates in to a single line.}
    \label{fig4}
\end{figure}

As the maximum speed of sound is reduced, the allowed range of possible $n(\muB)$ diminishes rapidly leading to tighter bounds on the EoS. This is demonstrated in \cref{fig4} showing the range of allowed values of $p(\epsilon)$ at different values of $\cslim$.  Interestingly, while the so-called conformal bound --- that is, $\cslim = 1/3$ --- is consistent with the high-density constraint, imposing this condition gives an extremely strong constraint and forces the EoS to have a specific shape at all densities.

Decreasing $\cslim$ even further,  we eventually close the gap between the lower and upper bounds completely, shown in \cref{fig4} as a red line. At this specific value of $\cslim = \csminlim$ the allowed region degenerates into a single line. This is the minimum limiting value for sound speed for which the low- and high- density limits can be connected in a consistent way. Thus $c_s^2 > \csminlim$ has to be reached by any EoS at some density. For $X=[1,4]$, we get $\csminlim = [0.30,0.32]$, just below the asymptotic conformal value of pQCD.

\section{Discussions}

Our results demonstrate the nontrivial and robust constraints on the EoS at NS densities arising from pQCD calculations. These constraints are obtained only when interpolating the full EoS $p(\muB)$ instead of its reduced form $p(\epsilon)$.

In order to demonstrate the practical utility of the results we have checked the consistency of the large number of publicly available EoSs \cite{Typel:2013rza, Oertel:2016bki} against new constraints; see \cref{figA2} in the Supplemental Material.

\begin{figure}
    \centering
\includegraphics[width=0.45\textwidth]{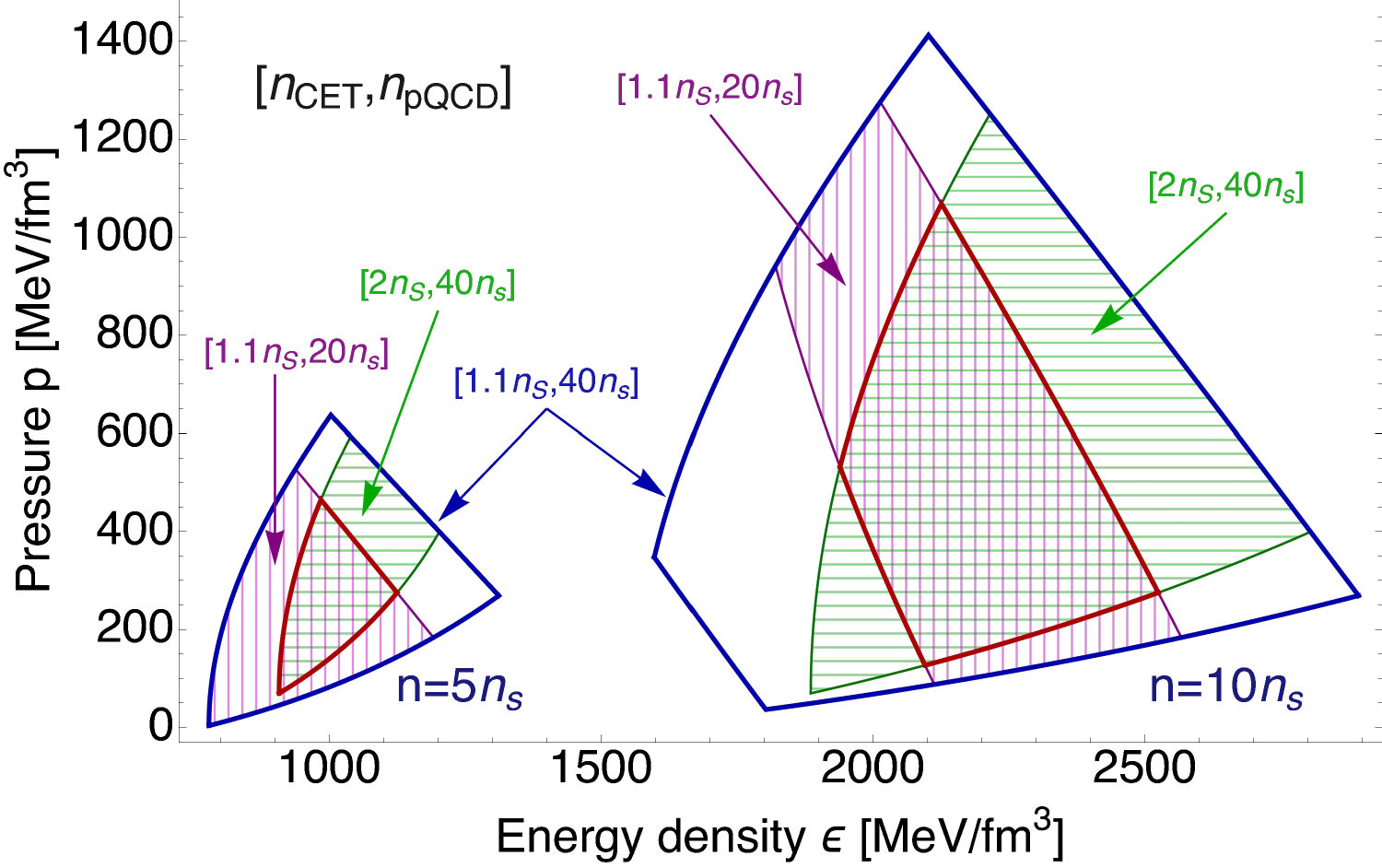}
    \caption{The expected impact of future calculations on the theoretical constraints to $\epsilon$ and $p$ at NS densities obtained by extrapolating the CET and pQCD calculations to intermediate densities. Numbers in square brackets show the end point of extrapolation of the CET and pQCD calculations.}
    \label{fig5}
\end{figure}

The results highlight the complementarity of the high- and low-density calculations. This is further demonstrated in \cref{fig5} depicting the expected effect of future calculations on the allowed region of $\epsilon-p$ -values at $5 n_s$ and $10 n_s$. We anticipate future results by extrapolating either CET ("stiff") up to $n=2 n_s$ using values tabulated in \cite{Hebeler:2013nza} and/or extrapolating the pQCD results ($X=2$) of \cite{Gorda:2021znl} down to $n=20 n_s$ (corresponding to $\muB = 2.1$ GeV). Improvements of this magnitude have been suggested, e.g., in \cite{Fujimoto:2020tjc, Fernandez:2021jfr}. We see that both the low- and the high-density calculations have a potential to significantly constrain the EoS in the future and that the greatest benefit is obtained by combining the both approaches given by the overlapping region in \cref{fig5}. These results highlight the importance of pursuing both the low- and high-density calculations in the future.

\section{Acknowledgments}
We thank Eemeli Annala, Tyler Gorda, Tore Kleppe, Kai Hebeler, Achim Schwenk, and Aleksi Vuorinen for useful discussions. We also thank Sanjay Reddy for posing questions on multiple occasions about the role of high-density constraints that in part motivated this paper. 

\appendix

\section{Boundaries on the $\epsilon - p$ -plane}
\label{AppA}
Utilizing the Euler equation
$
\epsilon = - p + \muB n,
$
 one can map bounds from $\muB - n$ -plane to the corresponding limits on the $\epsilon - p$ -plane.
 In order to find the extremal allowed values of $p(\epsilon)$, consider lines of fixed enthalpy $h = \epsilon + p = \muB n$. These lines correspond on one hand to diagonal lines $p(\epsilon) = -\epsilon + h$ on the $\epsilon - p$ -plane, and on the other hand to hyperbolas $n(\muB) = h / \muB$ on the $\muB - n$ -plane. Therefore, minimising or maximizing the pressure for a constant $h$ on the $\muB - n$ -plane gives the minimal and maximal pressure on the corresponding isenthalpic line on the $\epsilon - p$ -plane. 
 
 Substituting $n = h/\muB$ in \cref{eq:pmin} we readily observe that the smallest minimal pressure along isenthalpic line, $p_{\rm min}(\muB, h/\muB)$, is obtained at smallest value $\muB$ allowed by the constraints, that is, at the crossing of the isenthalpic line and $n_{\rm max}(\muB)$.

 Similarly, substituting  $n = h/\muB$ in \cref{eq:pmax1} and \cref{eq:pmax2} we observe that the maximal pressure $p_{\rm max}(\muB, h/\muB)$ obtains its largest value for the largest value $\muB$ allowed, at the crossing of the isenthalpic line and $n_{\rm min}(\muB)$. 
 
 Therefore the allowed range of values in the $\epsilon - p$ -plane is bounded from below by the line  $\{ \epsilon_{\rm max}(\mu) , p_{\rm min}(\muB, n_{\rm max}(\muB)) \}$ with $\muC < \muB < \muQ$, where
 \begin{align}
     &\epsilon_{\rm max}(\muB) 
     =  -p_{\rm min}(\muB, n_{\rm max}(\muB)) + \muB n_{\rm max}(\muB) \\
     & =  
     \left\{
     \begin{array}{cc}
     \frac{\left(\muB^2+\muQ^2\right) \left(\muB^2 \nC+\muC(2 \pC-\muC\nC)\right)-4 \muB^2 \muC\pQ}{2 \muC(\muB-\muQ) (\muB+\muQ)}, & \muB < \muc \\
      \frac{1}{2} ((\muB^2 \nQ)/\muQ + \muQ \nQ - 2 \pQ),  & \muB > \muc.
     \end{array}
    \right. \nonumber
 \end{align}
 This corresponds to the lower bound in \cref{fig1} and \cref{fig4}. 
 In \cref{fig3}, the line $\{\muB,n_{\rm max}(\muB),p_{\rm min}(\muB,n_{\rm max}(\muB))\}$ corresponds to the dashed green line marked $n_{\rm max}$. 
 
 Similarly the allowed $\epsilon - p$ values are bounded by above by $\{\epsilon_{\rm min}(\muB), p_{\rm max}(\muB, n_{\rm min}(\mu_B)) \}$, with
 \begin{align}
    & \epsilon_{\rm min}(\muB) 
      =  -p_{\rm max}(\muB, n_{\rm min}(\muB)) + \muB n_{\rm min}(\muB) \\
     & = 
     \left\{
     \begin{array}{cc}
    \frac{1}{2} ((\muB^2 \nC)/\muC + \muC \nC - 2 \pC)
     & \muB < \muc \\
     \frac{ 
   \frac{\muB^4}{\muC^2} \frac{\nQ}{\muQ} + 
 (\frac{\muB}{\muC})^2 (\muC^2 \frac{\nQ}{\muQ} - \muQ \nQ + 4 \pC - 2 \pQ) + 2 \pQ - \muQ \nQ 
    }{2 ((\frac{\muB}{\muC})^2 - 1)},  & \muB > \muc.
      \end{array}
    \right. \nonumber
 \end{align}

 This corresponds to the upper bound of pressure in \cref{fig1} and  \cref{fig4}. In \cref{fig3}, the line $\{\muB,n_{\rm min}(\muB),p_{\rm max}(\muB,n_{\rm min}(\muB))\}$ corresponds to the dashed green line marked $n_{\rm min}$.

Additionally, in \cref{figA1} we present the impact of the pQCD renormalization parameter variation on the allowed area for the fixed number density $5n_s$ and $10n_s$. It can be readily seen from the figure that for $5n_s$ pQCD input excludes 65$\%$ (for $X=1$) of the otherwise allowed area, and for $X=2$ only 25$\%$ of the $\epsilon - p$ values is allowed. The expressions for the areas without pQCD constraints are easily obtained by sending $\muc \rightarrow \infty$ and $n_c \rightarrow \infty$ and so taking the expressions corresponding to $\mu < \muc$ and $n< n_c$. 
 
\renewcommand{\thefigure}{A1}
\begin{figure}
    \centering
    \includegraphics[width=0.45\textwidth]{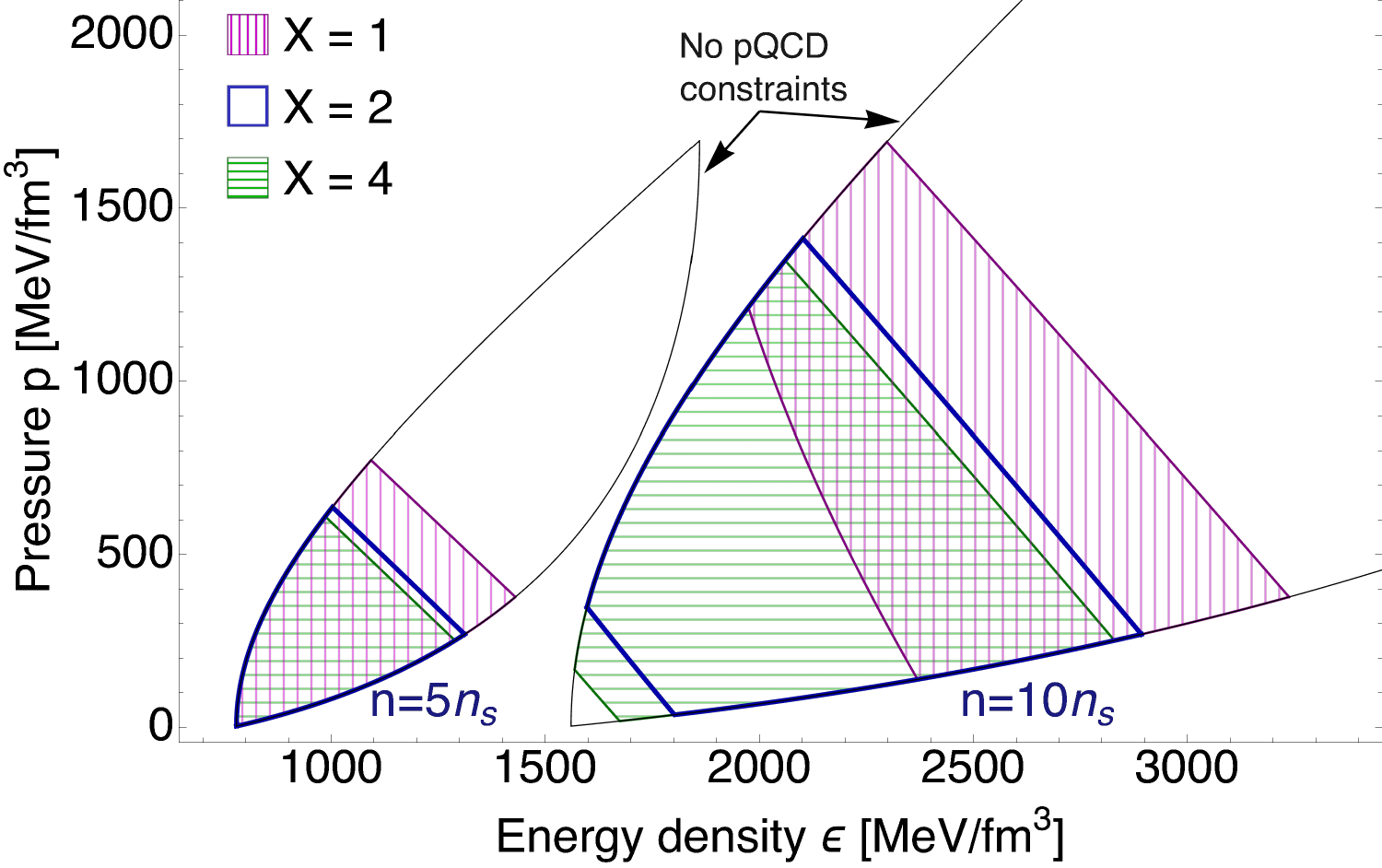}
    
    \caption{The impact of the pQCD renormalization parameter variation on the theoretical constraints to $\epsilon$ and $p$ at NS densities. Blue area denoted as $X = 2$ corresponds to blue area in \cref{fig1}. Purple and green areas correspond to allowed area for renormalization parameter $X=1$ and $X=4$, respectively. Thin black line corresponds to allowed area without high-density input.}
    \label{figA1}
    
\end{figure}

\section{Boundaries with arbitrary $\cslim$}
\label{AppB}

In this appendix we give generalizations of the bounds discussed in the main text for limiting speed of sound $\cslim=1$ to arbitrary $\cslim$. 

The line with the constant speed of sound is given by 
\begin{equation}
    n(\muB)= n_0(\muB/\mu_0)^{1/c_s^2}.
\end{equation}
The assumption that no EoS is allowed to be stiffer than this with the additionally imposed integral constraints leads to the minimal and maximal densities at fixed $\mu$
\begin{equation}
    n_{\rm min} = 
    \left\{
    \begin{array}{ll}
       \nC  \left(\frac{\muB}{\muC}\right)^{1/c_s^2}, \mu < \muc \\
       \frac{\left(\frac{\muB}{\muC}\right)^{1/c_s^2} \left(c_s^2\left(\muB \nQ \left(\frac{\muB}{\muQ}\right)^{1/c_s^2}-\muQ \nQ+\Delta p \right)+\Delta p\right)}{c_s^2\left(\muB \left(\frac{\muB}{\muC}\right)^{1/c_s^2}-\muC\right)}, \mu > \muc 
    \end{array}
    \right.
\end{equation}
and 
\begin{equation}
    n_{\rm max} = 
    \left\{
    \begin{array}{ll}
       \frac{\left(\frac{\muB}{\muQ}\right)^{1/c_s^2} \left(c_s^2 \left(\muB \nC \left(\frac{\muB}{\muC}\right)^{1/c_s^2}-\muC \nC-\Delta p\right)-\Delta p \right)}{c_s^2 \left(\muB \left(\frac{\muB}{\muQ}\right)^{1/c_s^2}-\muQ\right)}, \mu < \muc
       \\
          \nQ  \left(\frac{\muB}{\muQ}\right)^{1/c_s^2}, \mu > \muc 
    \end{array}
    \right.
\end{equation}
where 
\begin{equation}
\label{eq:muc_2}
    \muc = \left(\frac{\muQ^{1/c_s^2} (c_s^2 (\muC \nC-\muQ \nQ+\Delta p)+\Delta p)}
    {c_s^2 \left(\nC \left(\frac{\muQ}{\muC}\right)^{1/c_s^2}-\nQ\right)}\right)^{\frac{c_s^2}{c_s^2+1}}.
\end{equation}

For any given $\mu$ the largest allowed latent heat associated to a phase transition is given by
\begin{align}
    &\ \ \ Q(\muB) = \mu (n_{\rm max}(\muB)-n_{\rm min}(\muB)) \\
    &= 
    \frac{\muB \left(\frac{\muB}{\muC \muQ}\right)^{1/c_s^2} \left(c_s^2 \nC \muQ^{\frac{1}{c_s^2}+1}-\muC^{1/c_s^2} (c_s^2 (\muC \nC+\Delta p)+\Delta p)\right)}{c_s^2 \left(\muB \left(\frac{\muB}{\muQ}\right)^{1/c_s^2}-\muQ\right)} \nonumber
\end{align}
for $\mu < \muc$. 
For $\mu>\muc$ the expression is obtained by changing $H \leftrightarrow L$ in the above expression and multiplying the whole expression by -1. The phase transition with largest allowed latent heat takes place at $\mu=\muc$ and has the latent heat 
\begin{equation}
Q = \epsilon_{\rm H}-\epsilon_{\rm L} - \frac{\pQ-\pC}{c_s^2}.
\end{equation}

The line with largest pressure at fixed $\muB$ is given by
\begin{equation}
    n_c(\muB) = n_{\rm max}(\muC)\left( \frac{\muB}{\muC}\right)^{1/c_s^2}.
\end{equation}

The minimal and maximal pressures at given point $\{ \mu_0, n_0 \}$ generalize simply to
\begin{equation}
    p_{\rm min}(\mu_0,n_0) = \pC 
    + \frac{c_s^2  }{1+ c_s^2} 
    \left[
    \mu_0 
     -\muC \left( \frac{\muC}{\mu_0}\right)^{1/c_s^2}
    \right]  n_{\rm min}(\mu_0)
\end{equation}
and for $n_0 < n_c(\mu_0)$
\begin{equation}
    p_{\rm max}(\mu_0,n_0) = \pC 
    + \frac{c_s^2  }{1+ c_s^2} 
    \left[
    \mu_0 
     -\muC \left( \frac{\muC}{\mu_0}\right)^{1/c_s^2}
    \right]  n_0
\end{equation}
and for $n_0 > n_c(\mu_0)$
\begin{equation}
    p_{\rm max}(\mu_0,n_0) = \pQ
    + \frac{c_s^2  }{1+ c_s^2} 
    \left[
    \mu_0 
     -\muQ \left( \frac{\muQ}{\mu_0}\right)^{1/c_s^2}
    \right]  n_0.
\end{equation}

Reducing the $\cslim$, the allowed values of $n(\mu)$ eventually degenerate into a line. This line is either a line of constant speed of sound starting from $\{\muC, \nC\}$ with a phase transition at $\muQ$, or a line of constant speed of sound terminating to $\{\muQ, \nQ\}$ with a phase transition at $\muC$. Which of these cases is realised depends on the input parameter values, the criteria for it is given by 
\begin{align}
\label{eq:criteria}
 \frac{\pQ-\pC}{\epsilon_{\rm H}-\epsilon_{\rm L} } = \frac{\log(\muQ/\muC)}{\log(\nQ/\nC)} .
\end{align}

If the left hand side of the \cref{eq:criteria} is bigger than right hand side than the value for $\csminlim$ is given by the solution of the equation 
\begin{align}
\label{eq:csminlimA}
    \int_{\muC}^{\muQ} d \mu \, \nQ \left( \frac{\muB}{\muQ} \right)^{1/\csminlim} & = \Delta p.
\end{align}

Otherwise the equation reads 
\begin{align}
\label{eq:csminlimB}
    \int_{\muC}^{\muQ} d \mu \, \nC \left( \frac{\muB}{\muC} \right)^{1/\csminlim} &
   =  \Delta p.
\end{align}
In general, these equations need to be solved numerically. The another way of computing $\csminlim$ would be finding which of these cases happen first. Instead of using criteria \cref{eq:criteria} one can solve both eqs. (\ref{eq:csminlimA}),(\ref{eq:csminlimB}) and take maximum of the solutions in order to find $\csminlim$.

\section{Comparison to publicly available EoSs}
\label{AppC}

\Cref{figA2} visualises the practical utility of the results by comparing  a large number of model EoSs with the new constraints. The set of models contains all zero-temperature EoSs in $\beta$-equilibrium from the public CompOSE database \cite{Typel:2013rza, Oertel:2016bki}. The constraints are applied using the construction for the fixed number density, defined in eqs. (\ref{eq:pmin},\ref{eq:pmax1},\ref{eq:pmax2}). For every given point ($n > 1.1n_s$) of the EoS we use 3 values: pressure, energy density and number density. We find that a large portion of models
in the database violate the conditions within the range provided.

\renewcommand{\thefigure}{A2}
\begin{figure}[th!]
    \centering
    \includegraphics[width=0.45\textwidth]{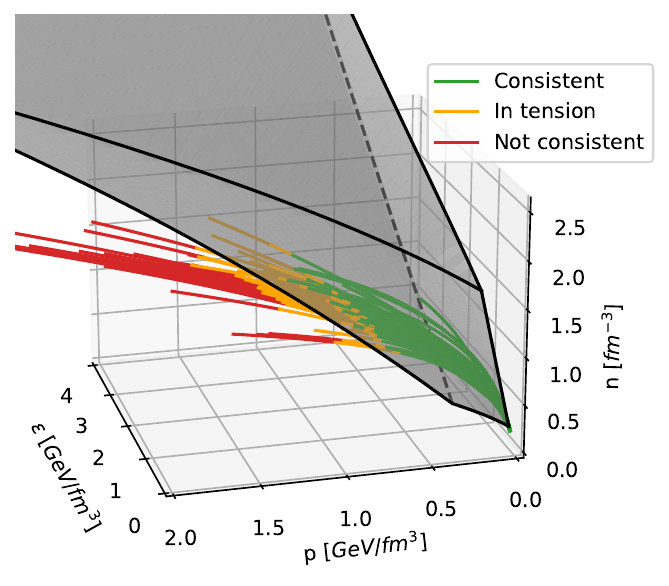}
    
    \caption{Consistency of all zero temperature EoSs in $\beta - $equilibrium from the CompOSE database with the new constraints. Green segment "Consistent" displays where EoS is consistent with the new constraints for any value of X in the range [1,4]. Yellow segment "In tension" represents region where EoS is consistent for some values of X in the range. Red segment "Not consistent" shows where EoS is not consistent with high-density input for any X = [1,4]. Gray area represents region allowed by the new constraints in the $\epsilon - p - n$ - space for X=2. }
    \label{figA2}
    
\end{figure}

\newpage

\bibliography{analytic.bib}

\end{document}